\documentclass[preprint,12pt]{elsarticle}

\usepackage{amssymb}
\usepackage{amsmath}
\usepackage{graphicx}
\usepackage{bm}
\usepackage{epsfig}
\usepackage{natbib}
\usepackage{mathcomp}

\newcounter{bla}

\journal{Computer Physics Communications}

\begin{document}

\begin{frontmatter}



\title{CHICOM: A code of tests for comparing unweighted and weighted histograms
  and two weighted histograms}


\author[a]{N.D. Gagunashvili \corref{author}}

\cortext[author] {Corresponding author.\\\textit{E-mail address:} nikolai@simnet.is}
\address[a]{University of Akureyri, Borgir, v/Nordursl\'od, IS-600 Akureyri, Iceland}

\begin{abstract}
A  Fortran-77 program for calculating test statistics to compare weighted histogram with an unweighted histogram  and  two histograms with weighted entries  is presented.  The code calculates test statistics for cases of histograms with normalized weights of events and  unnormalized weights of events.
\end{abstract}


\end{frontmatter}



{\bf PROGRAM SUMMARY}
\vspace *{0.5 cm}

\begin{small}
\noindent
{\em Program Title:} CHICOM                                        \\
{\em Journal Reference:}                                      \\
{\em Catalogue identifier:}                                   \\
{\em Licensing provisions:} none                                  \\
{\em Programming language:} Fortran-77                                  \\
{\em Computer:} Any Unix/Linux workstation or PC  with a Fortran-77 compiler.                                              \\
{\em Classification:} 4.13, 11.9, 16.4, 19.4                   \\
{\em External routines/libraries used:} FPLSOR (M103) \cite{cern} and BRENT \cite{lib}\\
{\em Nature of problem:} The program calculates  test statistics for comparing two weighted histograms and  an unweighted histogram with a weighted one.\\
{\em Solution method:}  Calculation of test statistics is done according  formulas presented in Ref. \cite{gagu_com}. \\
{\em Running time:} 0.001 sec for 5 bins histogram.\\

\end{small}

\section{Introduction}

  A histogram with $m$ bins for a given probability density function  $p(x)$ is used to estimate the probabilities $p_i$ that a random event  belongs in  bin $i$:
\begin{equation}
p_i=\int_{S_i}p(x)dx, \; i=1,\ldots ,m. \label{p1}
\end{equation}
 Integration in (\ref{p1}) is carried out over the bin  $S_i$ and $\sum_1^m p_i=1$.
A histogram can be obtained as a result of a random experiment with the probability
 density function $p(x)$.

A frequently used technique in data analysis is the comparison of two distributions through the comparison of histograms.
The  hypothesis of homogeneity  \cite{cramer} is that the two histograms
  represent random  values with  identical distributions.
  It is equivalent to there existing $m$ constants $p_1,...,p_m$,
 such that $\sum_{i=1}^{m} p_i=1$,
 and the probability  of  belonging  to the  $i$th bin for some  measured value
 in both experiments is  equal to $p_i$.

Let us denote the number of random events belonging to the $i$th bin of the first and second histograms as $n_{1i}$ and $n_{2i}$, respectively. The total number of events in the histograms are equal to $n_j=\sum_{i=1}^{m}{n_{ji}}$, where $j = 1, 2$.

As shown in \cite{cramer} the statistic
\begin{equation}
\frac{1}{n_1n_2} \sum_{i=1}^{m}{\frac{(n_2n_{1i}-n_1n_{2i})^2}{n_{1i}+n_{2i}}} \label{xsquar1}
\end{equation}
has approximately a $\chi^2_{m-1}$ distribution if hypothesis of homogeneity is valid.

Weighted histograms are often obtained as a result of Monte-Carlo
simulations.
References \cite{muon,weight,astro} are examples of research on high-energy physics, statistical mechanics, and astrophysics using such histograms.

To define a weighted histogram let us write the probability $p_i$
(\ref{p1}) for a given probability density function  $p(x)$  in the
form
\begin{equation}
p_i= \int_{S_i}p(x)dx = \int_{S_i}w(x)g(x)dx, \label{weightg}
\end{equation}
where
\begin{equation}
w(x)=p(x)/g(x) \label{fweight}
\end{equation}
 is the weight function and $g(x)$ is some other probability density function. The function $g(x)$ must be $>0$ for points $x$, where $p(x)\neq 0$. The weight $w(x)=0$ if $p(x)=0$, see Ref.  \cite{Sobol}.  Because of the condition $\sum_ip_i=1$ further we will call the above defined weights  normalized weights as opposed to the unnormalized weights $\check{w}(x)$ which are $\check{w}(x)=const\cdot w(x)$.

The histogram with normalized weights
 was obtained from a random experiment with a probability density function $g(x)$, and the weights of the events were calculated according to (\ref{fweight}). Let us denote the total sum of the weights of the events in the $i$th bin of the histogram with normalized weights as
\begin{equation}
W_i= \sum_{l=1}^{n_i}w_i(l), \label{ffweight}
\end{equation}
where $n_i$ is the number of events at bin $i$ and $w_i(l)$ is the weight of the $l$th event in the $i$th bin. The total number of events in the histogram is equal to $n=\sum_{i=1}^{m}{n_i}$, where $m$ is the number of bins. The quantity $\hat{p}_i= W_{i}/n$ is the estimator of $p_i$ with the expectation value $ E \,
[\hat{p_i}]=p_i$. Note that in the case where $g(x)=p(x)$, the weights of the events are equal to 1 and the histogram with normalized weights
is the usual histogram with unweighted entries.

Let us introduce notations need for the description of tests for comparing histograms:\\
\begin{itemize}
\item $W_{ji}=\sum_{l=1}^{n_{ji}}w_{ji}(l)$ -- the total sum of the weights of the events in the $i$th bin of the $j$th the histogram with normalized weights;\\
    \item    $r_{ji}=\sum_{l=1}^{n_{ji}}{w}_{ji}(l)/\sum_{l=1}^{n_{ji}}{w}_{ji}^2(l)$ -- estimator of the ratio of  moments  in the $i$th bin of the $j$th histogram with normalized weights.
        \end{itemize}
And the same quantities we introduce for the histograms with unnormalized weighted entries: \\
\begin{itemize}
\item $ \check{W}_{ji}=\sum_{l=1}^{n_{2i}}\check{w}_{ji}(l)$\\ \item     $\check{r}_{ji}=\sum_{l=1}^{n_{ji}}\check{{w}}_{ji}(l)/\sum_{l=1}^{n_{ji}}\check{{w}}_{ji}^2(l)$
\end{itemize}
Notice that $W_{ji}=n_{ji}$ and $r_{ji}=1$ for  histograms with unweighted entries.

Three types of statistics used for comparing   histograms are presented at Ref \cite{gagu_comp}.

\noindent
\textbf{Histograms with normalized weighted entries.}\\
Let us introduce the statistic
\begin{equation}
_1X_k^2=\sum_{j=1}^2\frac{1}{n_j} \sum_{i \neq k} \frac{ r_{ji}W_{ji}^2}{ p_{i}}+\sum_{j=1}^2\frac{1}{n_j}
\frac{(n_j-\sum_{i \neq k}r_{ji}W_{ji})^2}{1-\sum_{i \neq k}r_{ji}
 p_{i}}-\sum_{j=1}^2n_j .\label{stdd3}
\end{equation}
with the sums in (\ref{stdd3}) extending over all bins $i$ except one bin $k$. In
the equation  (\ref{stdd3}), the probabilities $p_i$ are unknown, and
estimators $\hat p_i$  of the probabilities are found by  minimization of
(\ref{stdd3}). We denote by $_1\hat X_k^2$ the value of $_1{X}_k^2$
after substitution of the estimators  $\hat p_i$ into
(\ref{stdd3}). As shown in \cite{gagu_comp}, the statistic
\begin{equation}
_1X^2= \textrm {Med}\, \{_1\hat X_1^2, \,  _1\hat X_2^2,  \ldots ,
\,_1\hat X_m^2\}\label{stdavu}
\end{equation}
approximately has a $\chi^2_{m-1}$  distribution if the hypothesis of
homogeneity is valid.

\noindent
\textbf{Histograms with  unnormalized weighted entries.}\\
Let us introduce the statistic
 \begin{equation}
{_2{X}}_k^2= \sum_{j=1}^2\frac{s_{kj}^2}{n_j}+2\sum_{j=1}^2 s_{kj},\label{stattwo1}
\end{equation}
 where
 \begin{equation}
 s_{kj}=\sqrt{\sum_{i \neq k}{\check{r}}_{ji} p_{i} \sum_{i \neq k}
{\check{r}}_{ji}\check{W}_{ji}^2/ p_{i}} - \sum_{i \neq
 k} {\check{r}}_{ji}\check{W}_{ji}.
 \end{equation}
 Again estimators  $\hat p_i$ of unknown probabilities $p_{i}$  are  found by  minimization of
(\ref{stattwo1}).
We denote by $_2\hat X_k^2$ the value of $_2{X}_k^2$
after substitution of the estimators  $\hat p_i$ into
(\ref{stattwo1}). As shown in \cite{gagu_comp}, the statistic
\begin{equation}
_2X^2= \textrm {Med}\, \{_2\hat X_1^2, \,  _2\hat X_2^2,  \ldots ,
\,_2\hat X_m^2\}\label{stdavu}
\end{equation}
approximately has a $\chi^2_{m-2}$  distribution if the hypothesis of
homogeneity is valid.

\noindent
\textbf{Histograms with normalized and unnormalized weighted entries.}\\
Let us introduce the statistic
\begin{equation}
 _3{X}_k^2=\frac{1}{n_1} \sum_{i \neq k} \frac{r_{1i}W_{1i}^2}{ p_{i}}+\frac{1}{n_1}
\frac{(n_1-\sum_{i \neq k}r_{1i}W_{1i})^2}{1-\sum_{i \neq k}r_{1i}
p_{i}}-n_1+ \frac{ s_{k2}^2}{n_2}+2s_{k2}. \label{sss}
\end{equation}
We denote by $_3\hat X_k^2$ the value of $_3{X}_k^2$
after substitution of the estimators  $\hat p_i$ into
(\ref{sss}). As shown in \cite{gagu_comp}, the statistic
\begin{equation}
_3X^2= \textrm {Med}\, \{_3\hat X_1^2, \,  _3\hat X_2^2,  \ldots ,
\,_3\hat X_m^2\}\label{stdavu}
\end{equation}
approximately has a $\chi^2_{m-2}$  distribution if the hypothesis of
homogeneity is valid.

The chi-square approximation is asymptotic. This means that the critical values may not be valid if the expected frequencies are too small. The use of the chi-square test is inappropriate
 if any expected frequency is $<1$, or if the expected frequency is $<5$ in $>20\%$ of the bins for either histogram. This restriction observed in the usual chi-square test \cite{moore} is quite reasonable for the proposed test.

 \textbf{Information for readers.} Recently, another paper dedicated to weighted histograms has been published in "Computer Physics Communication``, see Ref. \cite{chiwei}. The same author has presented a program for  goodness of fit test for histograms with weighted and unweighted entries. The test is used in a data analysis for comparison theoretical frequencies with frequencies represented by histogram.

\section{Computer program}

CHICOM is a subroutine which can be called from the Fortran programs for calculating test statistics $_1X^2$, $_2X^2$ and $_3X^2$.\\

\noindent
{\bf Usage}\\

\noindent
\verb"CALL CHICOM(AEX,ERAEX,NEV,AMC,ERAMC,NMC,NCHA,MODE,STAT,NDF,IFAIL)"\\

\noindent
{\bf {\it {Input Data}}}\\

\noindent
AEX -- one dimensional real array of first weighted histogram content \\

\noindent
ERAEX -- one dimensional real array  of histogram content for  entries of first histogram with squares of weights.\\

\noindent
NEV -- number of events in the first histogram $n_{1}$\\

\noindent
AMC -- one dimensional real array of second weighted histogram content \\

\noindent
ERAMC -- one dimensional real array  of histogram content for  entries of second histogram with squares of weights.\\

\noindent
NMC -- number of events in the second  histogram $n_{2}$\\

\noindent
NCHA -- number of bins $m$ \\

\noindent
  MODE --  equal 1 for both histograms with normalized weights,
   equal 2 for both histograms with unnormalized weights
   equal 3 for first histogram with normalized weights and the second with unnormalized weights\\

\noindent
{\bf {\it {Output data}}}\\

\noindent
STAT -- test statistic\\

\noindent
NDF -- number of degree of freedom $l $ of the $\chi^2_{l}$ distribution if hypothesis $H_0$ is true (will be $l=m-1$ or $l=m-2$)\\

\noindent
IFAIL -- will be $>0$ if calculation is not successful.

\section{Test run}

 We take a distribution:
\begin{equation}
p(x)\propto \frac{2}{(x-10)^2+1}+\frac{1}{(x-14)^2+1} \label{weight}
\end{equation}
 defined on the interval $[4,16]$ and representing two so-called Breit-Wigner
  peaks \cite{breit}.
  Three cases of the probability density function $g(x)$ are
  considered

\begin{equation}
g_1(x)=p(x)   \label{prc}
\end{equation}

\begin{equation}
g_2(x)=1/12  \label{flat}
\end{equation}

\begin{equation}
g_3(x)\propto\frac{2}{(x-9)^2+1}+\frac{2}{(x-15)^2+1} \label{real}
\end{equation}

Distribution $g_1(x)$ (\ref{prc}) results in  a  histogram with unweighted entries, while distribution $g_2(x)$ (\ref{flat}) is a uniform distribution
 on the interval [4, 16]. Distribution $g_3(x)$ (\ref{real}) has the same form of parametrization as $p(x)$ (\ref{weight}), but with different values for the parameters.

 Three cases were considered:\\

\noindent
 \begin{tabular}{|l|l|l|l|l|}
   \hline
   \multicolumn{1}{|c|}{} & \multicolumn{2}{|l|}{First histogram} & \multicolumn{2}{|l|}{Second histogram } \\
   \hline
   \textnumero &type of weight& weight &type of weight & weight \\
   \hline
   1& normalized &$p(x)/g_1(x)=1$& normalized &$p(x)/g_1(x)=1$   \\
   2& unnormalized & $0.5p(x)/g_2(x)$& unnormalized & $2p(x)/g_3(x)$   \\
   3& normalized &$p(x)/g_1(x)=1$  & unnormalized& $0.5p(x)/g_3(x)$  \\
   \hline
 \end{tabular}
  \vspace *{1cm}

  For each case histograms with 5 bins were created  by simulation 500 entries for first histogram and 1000 entries   for the second one.  The results of the calculations are presented below.\\

\textbf{Test 1 }

\verb"             INPUT"\\

\noindent
\verb"AEX      11.0000   58.0000  234.0000  102.0000   95.0000"\\
\verb"ERAEX    11.0000   58.0000  234.0000  102.0000   95.0000"\\
\verb"NEV     500"\\
\verb"AMC      30.0000  119.0000  439.0000  182.0000  230.0000"\\
\verb"ERAMC    30.0000  119.0000  439.0000  182.0000  230.0000"\\
\verb"NMC    1000"\\
\verb"NCHA      5"\\
\verb"MODE      1"\\

\verb"             OUTPUT"\\

\noindent
\verb"STAT      4.7391"   \hspace *{3cm}       (p-value =  0.3151)\\
\verb"NDF       4"\\
\verb"IFAIL     0"\\

\textbf{Test 2}

\noindent
\verb"               INPUT"\\

\noindent
\verb"AEX       9.3018   22.8871  122.0670   51.6786   46.2622"\\
\verb"ERAEX     0.8026    7.7173  142.7876   27.7087   28.5724"\\
\verb"NEV     500"\\
\verb"AMC      68.9455  213.5029  898.8528  397.7258  419.0171"\\
\verb"ERAMC   108.3022  229.3163 3697.7102 1455.0262  699.6888"\\
\verb"NMC    1000"\\
\verb"NCHA      5"\\
\verb"MODE      2"\\

\verb"             OUTPUT"\\

\noindent
\verb"STAT      1.9111"   \hspace *{3cm}       (p-value =  0.5911)\\
\verb"NDF       3"\\
\verb"IFAIL     0"\\

\textbf{Test 3 }

\verb"             INPUT"\\

\noindent
\verb"AEX      17.0000   53.0000  225.0000  101.0000  104.0000"\\
\verb"ERAEX    17.0000   53.0000  225.0000  101.0000  104.0000"\\
\verb"NEV     500"\\
\verb"AMC      14.2303   53.9921  204.9794  111.6337  101.1128"\\
\verb"ERAMC     5.4897   14.5935  198.6223  103.7259   40.9275"\\
\verb"NMC    1000"\\
\verb"NCHA      5"\\
\verb"MODE      3"\\

\verb"             OUTPUT"\\

\noindent
\verb"STAT      1.4431"   \hspace *{3cm}       (p-value =  0.6955)\\
\verb"NDF       3"\\
\verb"IFAIL     0"\\

\end{document}